# Formation of ordered and disordered interfacial films in immiscible metal alloys

Zhiliang Pan[a], Timothy J. Rupert[a,b,*]

[a] Department of Mechanical and Aerospace Engineering, University of California, Irvine, California, 92697, USA

[b] Department of Chemical Engineering and Materials Science, University of California, Irvine, California, 92697, USA

[*] Corresponding Author. Tel.: 949-824-4937; e-mail: trupert@uci.edu

## Abstract

Atomistic simulations are used to study segregation-induced intergranular film formation in Cu-Zr and Cu-Nb alloys. While Cu-Zr forms structurally disordered or amorphous films, ordered films comprised of a second phase usually precipitate in Cu-Nb, with a critical nucleation size of ~1 nm below which the ordered phase cannot form. While the ordered film is retained at high temperature for a low energy $\Sigma11$ (113) boundary, a disordering transition is observed for a high energy $\Sigma5$ (310) boundary at low dopant concentrations. Finally, the effect of free surfaces on dopant segregation and intergranular film formation is investigated for both alloys.

Many important material properties such as fracture and corrosion are dominated by interfacial phenomena [1], meaning that control of interface structure would be extremely advantageous for better materials design. Complexions are interfacial features that are in thermodynamic equilibrium with their abutting phases and which have a stable, finite thickness [2]. Complexions and transitions between different complexion types have been shown to be responsible for such diverse behavior as abnormal grain growth [3], solid-state activated sintering [4], liquid-metal embrittlement in Ni-Bi [5], and a dramatic increase in the ductility of nanocrystalline Cu-Zr [6]. Complexions can be classified in many ways, but a popular formulation is the Dillon-Harmer complexion type [3] where higher numbers are assigned to thicker complexions which generally have increased levels of adsorbate. Dopants in monolayer, bilayer, or trilayer complexions form specific segregation patterns at the grain boundary (GB), while nanoscale intergranular films and wetting phases can be either structurally and chemically ordered or disordered [2]. GB pre-melting [7, 8] and amorphous intergranular films (AIFs) [3, 9] are examples of higher level complexions which exhibit both chemical and structural disorder.

Since dopant segregation and GB transitions are nanoscale phenomena, atomistic simulations are ideal tools for studying complexion formation. For example, molecular dynamics (MD) have been used to study the structure of calcium silicate intergranular films in silicon nitride crystals [10]. Atomistic Monte Carlo (MC) methods have also been used to study the pre-melting behavior of GBs in Cu-Ag alloys [11]. Frolov et al. used a hybrid MC/MD method to discover two different ordered complexions that can exist at a Σ5 (210) symmetric tilt GB in Cu-Ag [12]. The transition between these two complexion types occurred when the dopant concentration was increased for a fixed temperature. Frolov et al. also showed that variation of the number of atoms in the grain boundary plane, provided by the addition of a free surface in an MD simulation, could drive a structural transition even in GBs of pure Cu [13]. Recently, Pan and Rupert used atomistic simulations to study disordering transitions in Cu-Zr [9], finding that different ordered complexions first form, with the Dillon-Harmer type depending on the boundary's starting structure, and then a transition to a structurally disordered state occurs at high interfacial concentrations of Zr. Purely thermodynamic models for complexion formation have also been developed. For example, Luo and Shi performed a study on disordering in binary alloys, showing that equilibrium complexion diagrams can be created to see the effects of temperature and global dopant concentration on complexion formation [14]. Frolov and Mishin also developed a generalized thermodynamic theory which showed that up to four interface phases can be supported in a binary single-phase system [15]. However, the range of structural and chemical order that can be accessed in intergranular films is not well understood. All of the examples listed above which treated intergranular films describe disordered or amorphous complexions. It is not clear if that is a foregone conclusion, with the abutting crystal frustrating crystallization of the intergranular film for nanoscale thicknesses, or if this observation is merely a function of the alloys that have been studied to date. There is also no clear understanding of the competition between dopant segregation and complexion formation on GBs and surfaces.

In this paper, hybrid atomistic MC/MD simulations were used to study GB doping and complexion transitions in Cu-Zr and Cu-Nb alloys, in order to develop a better understanding of nanoscale intergranular film formation. These binary systems are both immiscible combinations, where the dopant wants to segregate to interfacial regions, but they differ in other thermodynamic features, such as heat of mixing and the existence of stable intermetallic phases. Specifically, Cu-Nb has a positive enthalpy of mixing [16] and is a poor glass former, so it should be difficult for disordered complexions to form in this system. We find that the Cu-Zr system only forms amorphous films whereas the Cu-Nb system usually forms ordered intergranular films through heterogeneous nucleation. However, at higher temperatures, a disordering transition is observed for Cu-Nb under certain conditions. The competing effect of free surfaces is also studied, showing that surface segregation and structural transition also occurs for Cu-Zr.

Hybrid MC/MD simulations were performed using the Large-scale Atomic/Molecular Massively Parallel Simulator (LAMMPS) code [17], with all MD simulations using a 1 fs integration time step. The interatomic potential for Cu-Zr system was developed by Mendelev et al. [18], with Cu-Cu and Zr-Zr interactions described using embedded-atom method (EAM) potentials and Cu-Zr interactions with a Finnis-Sinclair formulation. The Cu-Nb interactions were described using an EAM potential that is able to rebuild the main features of the experimental phase diagram in both solid and liquid states [16]. Bicrystals with two different tilt GBs were used as starting configurations so that the effect of GB character can be studied. One sample contains two $\Sigma 11$ (113) GBs, with GB energy of 0.022 eV/Å$^2$. The other sample contains two $\Sigma 5$ (310) GBs, with GB energy of 0.062 eV/Å$^2$. Each simulation cell is about 20 nm long, 20 nm tall, and 4–5 nm thick, with each containing ~200000 atoms. Periodic boundary conditions were applied in all three directions for the majority of simulations, but only in the length and thickness direction when studying the effect of free surfaces.

The samples were first equilibrated with a conjugate gradient minimization technique to the minimum potential energy state and then further relaxed with a Nose-Hoover thermos/barostat for 20 ps under zero pressure at 600 K and 1200 K. The doping with Zr or Nb solutes was then simulated using the MC method in a variance-constrained semigrand canonical ensemble [19] after every 100 MD steps, with the global composition of dopants fixed to different values. In this MC scheme, the chemical potential difference was given an initial guess and then adjusted during the simulation to achieve the required composition. However, the final composition can slightly deviate from the target value, with the deviation depending on how much the initial guessed chemical potential difference deviates from the authentic one for the simulated system. The acceptance probability and thus the speed of MC simulation also decreases due to this mismatch [19]. To solve this problem, the initial chemical potential difference was also adjusted every 10 MC steps based on the difference between the achieved and target global compositions. The initial chemical potential difference is increased if the achieved composition is lower than the target value and decreased otherwise. With this modified MC scheme, the authentic chemical potential difference can also be achieved, which is, in fact, the guessed value at the converged MC step. Here the MC/MD simulation is considered to be converged if the fitted slope of the system potential energy over the last 4000 MC steps is less than 0.001 eV/step. After equilibrium, another conjugate-gradient minimization technique was used to remove thermal noise in the grain interior and at the same time preserve the interfacial structure obtained during the doping process. Atomic configurations are visualized using an open-source visualization tool OVITO [20], with the local structure identified using adaptive common neighbor analysis (CNA) [21, 22]. Face centered-cubic (FCC) atoms are colored green, hexagonal close packed atoms red, body-centered cubic (BCC) atoms blue, icosahedral atoms yellow, and other atoms white. The volume of each atom is also calculated using Voronoi tessellation method to measure the average GB thickness.

Fig. 1 shows doping of the Cu-Zr and Cu-Nb alloys for the $\Sigma 11$ boundary at 600 K, which is only ~48% and ~44%, respectively, of the solidus temperature for binary alloys in each system [23, 24]. The frame on the left of each figure part shows the local structure of each atom, while the frames on the right in each part show the chemical distribution. GB segregation of the dopant species occurs in both systems, as shown in Fig. 1(a) and (d), yet there is a clear difference between the Nb and Zr induced structural transitions at the GB. The Cu-Zr alloy forms AIFs, as shown by the prevalence of atoms which could not be assigned a common crystal structure in Fig. 1(b)-(c). This is a typical segregation-induced GB premelting or transition from an ordered complexion to a disordered one, as extensively discussed in the literature [2, 25-27]. Increased dopant concentration leads to a thickening of this disordered region, consistent with prior reports of both Cu-Zr [9] and Cu-Ag [11] alloys. Cu-Nb, on the other hand, forms an ordered, crystalline phase at the GB through heterogeneous nucleation. It is essential to note that this phase is not a complexion, but rather a bulk phase that happens to nucleate at the grain boundary site. These interphases are entirely comprised of Nb atoms and have a BCC crystal structure but do not fully wet the boundary at first. Rather, small precipitates form with roughly ellipsoidal cross-sections. These intergranular precipitates are simply the pure Nb phase, which always nucleate at the GB here because it is a favored nucleation site due to its high relative energy compared to the crystal. As dopant composition is increased, Nb atoms prefer to segregate to the GB areas that will lengthen the ordered Nb phase. At higher Nb concentrations, the ordered film has spread all the way across the boundary and the second phase then begins to thicken.

Fig. 2 represents a quantitative perspective of the GB doping phenomena described above. Average GB thickness, calculated by dividing the volume of non-FCC atoms by twice the cross-sectional area of the computational cell, is shown as a function of global dopant composition in Fig. 2(a). The Cu-Zr system smoothly transforms into an AIF after ~1 at.% Zr is added, with thickness increasing as more Zr is added. The Cu-Nb transition to an ordered phase is shown in Fig. 1(e) and also the inset to Fig. 2(a). At ~0.6% global composition, the Nb dopants aggregate together in the boundary and form the first precipitate, resulting in a jump in average boundary thickness. It is important to note that this value is averaged over the entire boundary area (i.e., it includes the areas with and without second phases), so the average GB thickness is smaller than the minimum precipitate size. The minimum thickness of the smallest precipitate is estimated to be ~1 nm, as measured in normal to the GB plane. The average thickness of the boundary increases linearly with increasing composition after this nucleation event, with all of the new Nb segregating directly to the boundary and growing the BCC phase until the boundary is covered. Fig. 2(b) presents the chemical composition of the GB regions for the Cu-Zr and Cu-Nb systems. The composition of the Cu-Zr GB slowly increases with the thickening of the nanoscale AIF. Eventually, this composition would saturate at the liquidus composition when a wetting film is created [9], but this does not occur yet here. The calculated GB composition in the Cu-Nb system increases more quickly, but it is important to recall the structure of this ordered intergranular phase. The ordered film has a BCC structure and is comprised entirely of Nb atoms, yet the interface between the film and the Cu crystal interiors is also identified as part of the GB and these are Cu atoms. Hence, the calculated composition really demonstrates that the Nb-rich film is getting larger. Fig. 2(c) similarly presents the composition of the grain interior. In the case of Zr, the grain composition increases until the amorphous complexion is nucleated. After this point, the grain composition drops to a lower, constant level. The Nb concentration in the grain similarly increases at first, but then drops to a very small number extremely close to zero after the heterogeneous nucleation of the ordered BCC phase at the boundary. This behavior is related to the minimum size of Nb precipitates. The Cu lattice does not want to accept substitutional Nb solutes, but the creation of a second BCC phase would require overcoming a interface energy penalty for creating a new interface between the phases (or possibly a large strain energy penalty to maintain compatibility at the precipitate-grain interface if precipitates were coherent) and the grain interior becomes oversaturated with Nb. At some point, this competition between Nb supersaturation and interfacial energy shifts and a second phase can finally form.

While an ordered intergranular film is formed for Cu-Nb through heterogeneous nucleation rather than a complexion transition in Figs. 1 and 2, these simulations were performed at the relatively low temperature of 600 K. To find if these ordered films persist at high temperatures, similar doping simulations were run at 1200 K, or 88% of the solidus temperature where the alloy would begin to melt. Fig. 3(a) and (b) presents atomic snapshots of the doping process for the same $\Sigma11$ GB described above. Even up to a concentration of 3 at. % Nb, the boundary remains untransformed with no second phases present. The nucleation of the intergranular phase occurred well before this composition at the lower temperature, showing that high temperature delays the nucleation event. The right frame of Fig. 3(a) shows the chemical distribution within the system for the same sample. While the GB region is still enriched with Nb, it is obvious that the crystal interior can also accept a significant Nb concentration at this higher temperature, explaining the delayed nucleation. The ordered film does eventually form, albeit at a higher global Nb composition, as shown in Fig. 3(b). Recent work has also shown that GB character plays an important role in determining the likelihood of forming disordered intergranular films [9], so it is important to understand if this nucleation of an ordered second phase persists for other boundary types. Since high GB energy is roughly related to the ease of AIF formation, we repeat these same high-temperature simulations with a $\Sigma5$ (013) boundary. At low dopant composition in Fig. 3(c), the boundary thickens and is structurally disordered. No common crystalline structure can be assigned to the GB film, meaning it is an AIF. The chemical distribution shows that segregation occurs but some Nb clustering is also apparent. This result highlights that AIFs can form under certain conditions, namely high temperatures and at high energy boundaries, even for a system like

Cu-Nb which usually prefers a two-phase structure with ordered interfacial films. However, at higher Nb concentrations (Fig. 3(d)), this AIF transforms into an ordered phase again, suggesting that only a limited window for AIF formation exists in this alloy system.

Finally, we investigated the effect of free surfaces on intergranular film formation. Free surfaces can also be considered as defects, with higher energies that can induce dopant segregation and possibly complexion formation [28, 29]. Fig. 4 shows the dopant segregation behavior during selected compositions of the initial complexion or precipitate formation, intermediate progression, and final covering of the GB at 600 K. In both cases, the same type of film forms (AIF for Cu-Zr and ordered BCC film for Cu-Nb), with an AIF also found on the surface for Cu-Zr. For both systems, the dopants begin to aggregate at the triple junctions, likely because the GB energy is not balanced in the vertical direction, making the triple junction line unstable. Structural transition in this region can generate new interfaces between the matrix and the transformed structure that balance the GB energy, as confirmed by the wedge-shaped transformed region in the triple junction. As the Zr composition is increased in Figs. 4(b) and (c), the disordered region at the triple junctions grows to cover the free surface and the GBs, and then begins to thicken in both regions. The $\Sigma 5$ (013) GB shows the same segregation and transformation behavior (not presented here), indicating that free surfaces can compete for dopant segregation and complexion transitions in some alloy systems.

In summary, hybrid MC/MD simulations were used to study ordered and disordered nanoscale film formation in Cu-Nb and Cu-Zr while varying GB character and temperature. Cu-Zr forms amorphous nanoscale films at the GBs through a complexion transition, while Cu-Nb forms crystalline nanoscale precipitates through heterogeneous nucleation and eventually full intergranular films at GB sites. Near the melting temperature of Cu, however, disordered intergranular films can form at certain boundaries in Cu-Nb even though this system is a poor glass former. Free surfaces can compete with the GBs as potential segregation sites, depending on the dopant type. As a whole, the results presented here show that efficient glass-forming binary systems are more likely to form disordered complexions, while low solubility systems with positive enthalpies of mixing will usually form crystalline intergranular films through heterogeneous nucleation.

This research was supported by the U.S. Department of Energy, Office of Science, Basic Energy Sciences, under Award No. DE-SC0014232.

# References

[1] S. Dillon, M. Harmer, J. Luo, Grain boundary complexions in ceramics and metals: An overview, JOM 61(12) (2009) 38-44.

[2] P.R. Cantwell, M. Tang, S.J. Dillon, J. Luo, G.S. Rohrer, M.P. Harmer, Grain boundary complexions, Acta Mater. 62 (2014) 1-48.

[3] S.J. Dillon, M. Tang, W.C. Carter, M.P. Harmer, Complexion: A new concept for kinetic engineering in materials science, Acta Mater. 55(18) (2007) 6208-6218

[4] J. Luo, H. Wang, Y.-M. Chiang, Origin of Solid-State Activated Sintering in $Bi_2O_3$-Doped ZnO, Journal of the American Ceramic Society 82(4) (1999) 916-920.

[5] J. Luo, H. Cheng, K.M. Asl, C.J. Kiely, M.P. Harmer, The Role of a Bilayer Interfacial Phase on Liquid Metal Embrittlement, Science 333(6050) (2011) 1730-1733.

[6] A. Khalajhedayati, Z. Pan, T.J. Rupert, Manipulating the interfacial structure of nanomaterials to achieve a unique combination of strength and ductility, Nature Communications 7 (2016) 10802.

[7] J. Luo, V.K. Gupta, D.H. Yoon, H.M. Meyer, Segregation-induced grain boundary premelting in nickel-doped tungsten, Appl. Phys. Lett 87(23) (2005) 231902.

[8] J. Luo, Liquid-like interface complexion: From activated sintering to grain boundary diagrams, Current Opinion in Solid State and Materials Science 12(5–6) (2008) 81-88.

[9] Z. Pan, T.J. Rupert, Effect of grain boundary character on segregation-induced structural transitions, Phys. Rev. B 93(13) (2016) 134113.

[10] S.H. Garofalini, W. Luo, Molecular Dynamics Simulations of Calcium Silicate Intergranular Films between Silicon Nitride Crystals, Journal of the American Ceramic Society 86(10) (2003) 1741-1752.

[11] P.L. Williams, Y. Mishin, Thermodynamics of grain boundary premelting in alloys. II. Atomistic simulation, Acta Mater. 57(13) (2009) 3786-3794.

[12] T. Frolov, M. Asta, Y. Mishin, Segregation-induced phase transformations in grain boundaries, Phys. Rev. B 92(2) (2015) 020103.

[13] T. Frolov, D.L. Olmsted, M. Asta, Y. Mishin, Structural phase transformations in metallic grain boundaries, Nature Communications 4 (2013) 1899.

[14] J. Luo, X. Shi, Grain boundary disordering in binary alloys, Appl. Phys. Lett 92(10) (2008) 101901.

[15] T. Frolov, Y. Mishin, Phases, phase equilibria, and phase rules in low-dimensional systems, The Journal of Chemical Physics 143(4) (2015) 044706.

[16] L. Zhang, E. Martinez, A. Caro, X.-Y. Liu, M.J. Demkowicz, Liquid-phase thermodynamics and structures in the Cu–Nb binary system, Modelling and Simulation in Materials Science and Engineering 21(2) (2013) 025005.

[17] S. Plimpton, Fast Parallel Algorithms for Short-Range Molecular Dynamics, Journal of Computational Physics 117(1) (1995) 1-19.

[18] M.I. Mendelev, M.J. Kramer, R.T. Ott, D.J. Sordelet, D. Yagodin, P. Popel, Development of suitable interatomic potentials for simulation of liquid and amorphous Cu–Zr alloys, Philosophical Magazine 89(11) (2009) 967-987.

[19] B. Sadigh, P. Erhart, A. Stukowski, A. Caro, E. Martinez, L. Zepeda-Ruiz, Scalable parallel Monte Carlo algorithm for atomistic simulations of precipitation in alloys, Phys. Rev. B 85(18) (2012) 184203.

[20] A. Stukowski, Visualization and analysis of atomistic simulation data with OVITO–the Open Visualization Tool, Modelling and Simulation in Materials Science and Engineering 18(1) (2010) 015012.

[21] J.D. Honeycutt, H.C. Andersen, Molecular dynamics study of melting and freezing of small Lennard-Jones clusters, J. Phys. Chem. 91(19) (1987) 4950-4963.

[22] A. Stukowski, Structure identification methods for atomistic simulations of crystalline materials, Modelling and Simulation in Materials Science and Engineering 20(4) (2012) 045021.

[23] D. Arias, J.P. Abriata, Cu-Zr (Copper-Zirconium), Bulletin of Alloy Phase Diagrams 11(5) (1990) 452-459.

[24] H. Okamoto, Cu-Nb (Copper-Niobium), Journal of Phase Equilibria and Diffusion 33(4) (2012) 344.

[25] J. Luo, Grain boundary complexions: The interplay of premelting, prewetting, and multilayer adsorption, Appl. Phys. Lett 95(7) (2009) 071911.

[26] Y. Mishin, W.J. Boettinger, J.A. Warren, G.B. McFadden, Thermodynamics of grain boundary premelting in alloys. I. Phase-field modeling, Acta Mater. 57(13) (2009) 3771-3785.

[27] M. Tang, W.C. Carter, R. Cannon, Grain boundary order-disorder transitions, J. Mater. Sci. 41(23) (2006) 7691-7695.

[28] H. Hakkinen, U. Landman, Superheating, melting, and annealing of copper surfaces, Phys. Rev. Lett. 71(7) (1993) 1023-1026.

[29] J. Li, G. Wang, G. Zhou, Surface segregation phenomena in extended and nanoparticle surfaces of Cu–Au alloys, Surface Science 649 (2016) 39-45.

**Figures and Captions**

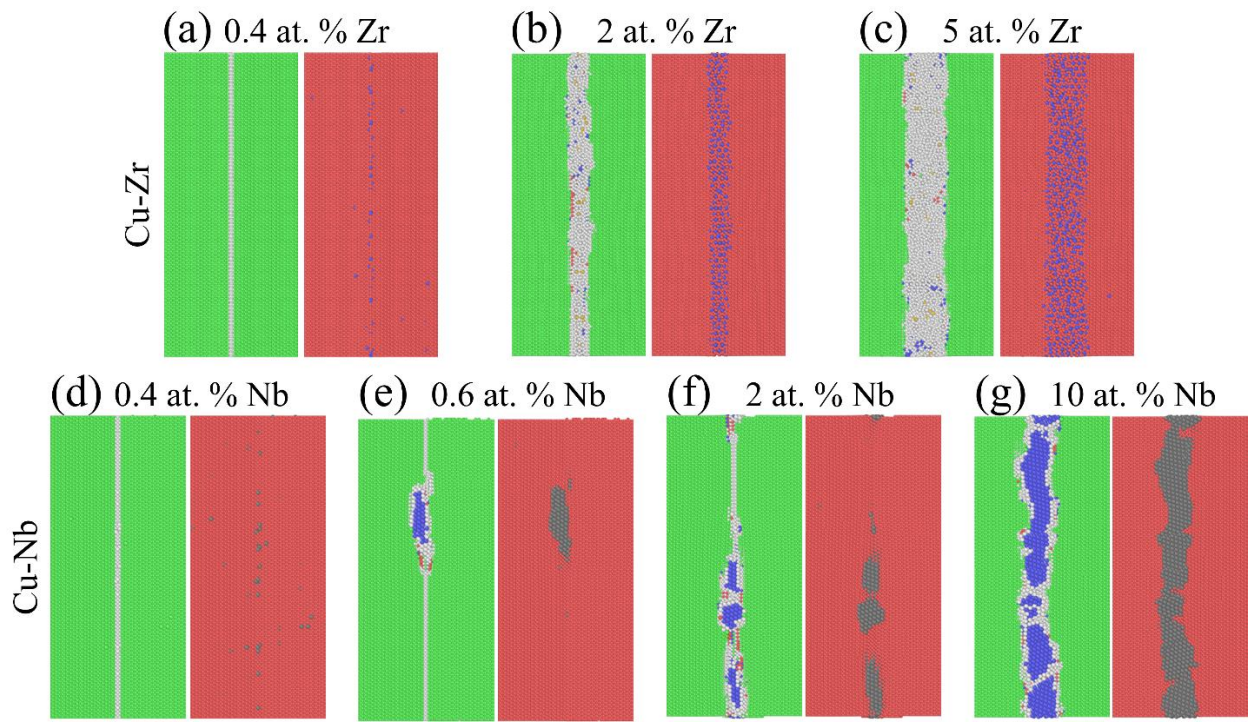

Fig. 1. Atomic snapshots of one of the two Σ11 (113) GBs in the sample, doped at 600 K in (a)-(c) the Cu-Zr and (d)-(g) Cu-Nb systems with periodic boundary conditions. The left frame in each pair shows the local structure of the atoms colored according to CNA. The right frame shows the chemistry of the boundary where peach atoms are Cu, blue atoms are Zr, and gray atoms are Nb.

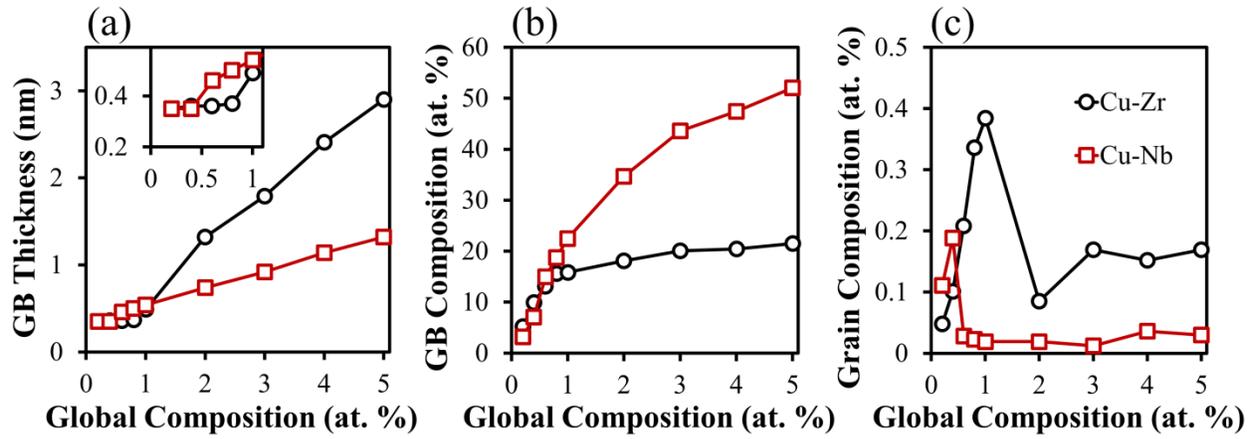

Fig. 2. (a) GB thickness, (b) GB composition, and (c) grain composition as a function of global dopant composition for the Σ11 (113) samples with periodic boundary conditions doped at 600 K.

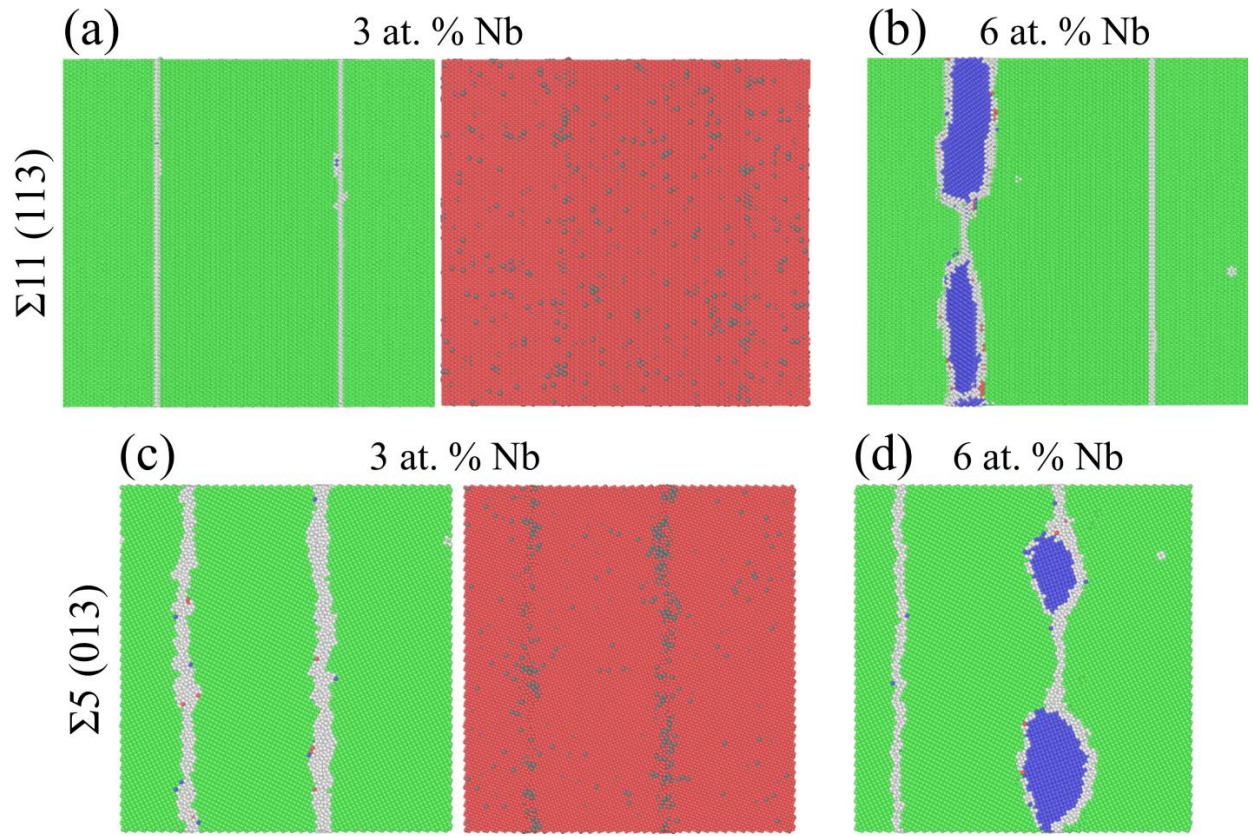

Fig. 3. Structural analysis of the Nb-doped (a), (b) Σ11 (113) boundary and (c), (d) Σ5 (013) boundary at 1200 K with periodic boundary conditions in all directions. The right frames in (a) and (c) show the chemical distribution of the Σ11 (113) boundary and Σ5 (013) boundary, respectively, where peach atoms are Cu and gray atoms are Nb.

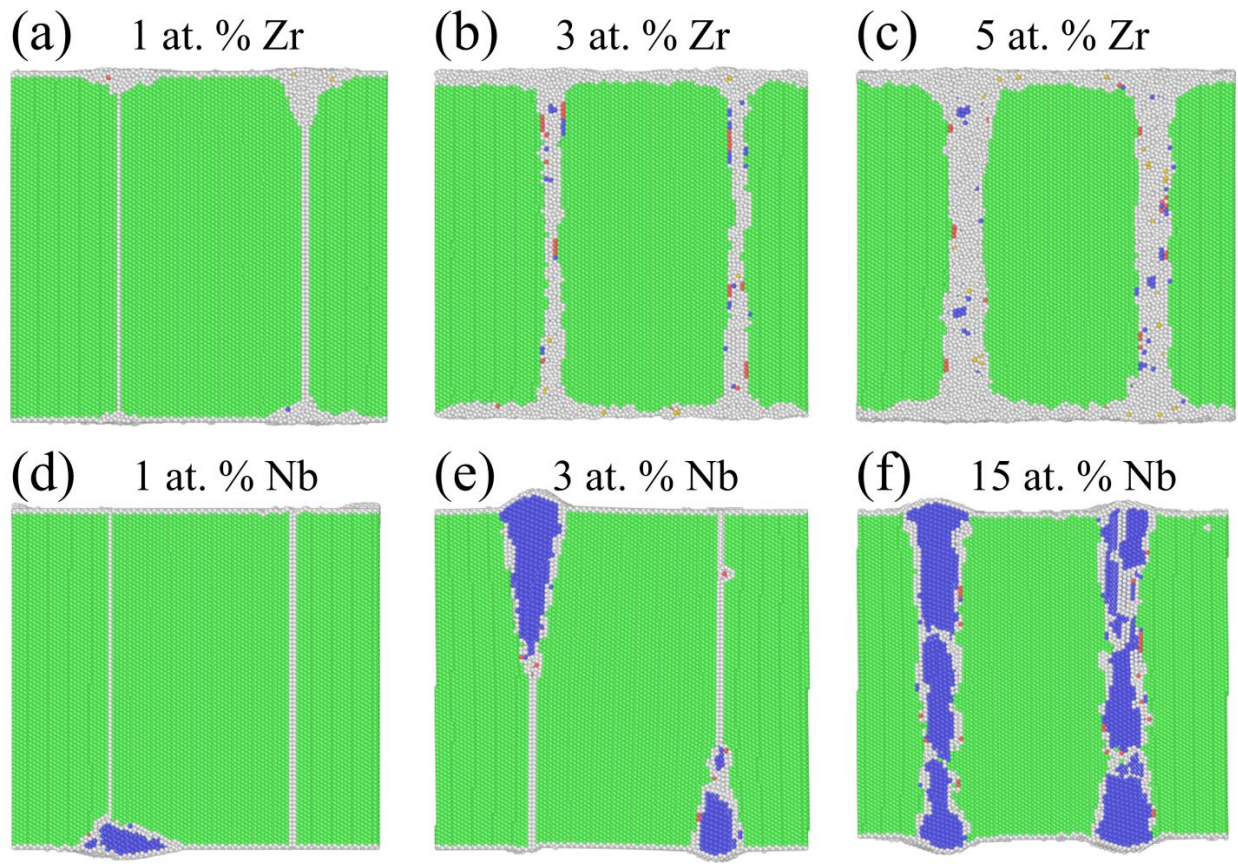

Fig. 4. Structural analysis of the Σ11 (113) samples with free surfaces at 600 K doped with (a)-(c) Zr and (d)-(f) Nb. Surface films are formed only in Cu-Zr.